\documentclass[aps,pra,twocolumn,showpacs,showkeys,superscriptaddress,reprint,floatfix]{revtex4-1} 
\usepackage{graphics} 
\usepackage{graphicx} 
\usepackage{amsmath} 
\usepackage{amssymb}
\usepackage{color}
\usepackage[caption=false]{subfig}
\begin{document} 

\title{Mixture of scalar bosons and two-color fermions in one dimension: Superfluid-insulator transitions} 
\author{R. Avella}
\affiliation{Departamento de F\'{\i}sica, Universidad Nacional de Colombia, A. A. 5997 Bogot\'a, Colombia.}
\author{J. J. Mendoza-Arenas}
\affiliation{Departamento de F\'{\i}sica, Universidad de los Andes, A. A. 4976 Bogot\'a, Colombia.}
\author{R. Franco} 
\affiliation{Departamento de F\'{\i}sica, Universidad Nacional de Colombia, A. A. 5997 Bogot\'a, Colombia.}
\author{J. Silva-Valencia} 
\email{jsilvav@unal.edu.co} 
\affiliation{Departamento de F\'{\i}sica, Universidad Nacional de Colombia, A. A. 5997 Bogot\'a, Colombia.} 

\date{\today} 

\begin{abstract}
Superfluid-insulator transitions in a one-dimensional mixture of two-color fermions and scalar bosons are studied within the framework of the Bose-Fermi-Hubbard model. Zero-temperature phase diagrams are constructed for repulsive intraspecies interactions and attractive or repulsive interspecies couplings. In addition to the trivial Mott insulator phases, we report the emergence of new non-trivial insulator phases that depend on the sign of the boson-fermion interaction. These non-trivial insulator phases satisfy the conditions $\rho_B\pm\rho_F=n$ and $\rho_B\pm \tfrac{1}{2}\rho_F=n$, with the plus (minus) sign for repulsive (attractive) interactions and $n$ an integer. Far from fermionic half-filling, the boson-fermion interaction drives a gapless-gapped transition in the spin sector. Our findings could be observed experimentally in state-of-the-art cold-atom setups. 
\end{abstract} 


\maketitle 

\section{\label{sec1}Introduction}
Rapid advances in the cold-atom field have allowed the observation of several predicted physical phenomena and have opened the possibility of experimenting with several dream scenarios~\cite{IBloch-RMP08,Esslinger-AR10,IBloch-NP12,Gross-S17}. One of the latter corresponds to mixtures of particles that obey Bose-Einstein or Fermi-Dirac statistics. Since the beginning of this century, experimentalists have mixed carriers with different statistics, using isotopes of different atoms or of the same type of atom~\cite{Truscott-S01,Schreck-PRL01,Hadzibabic-PRL02,Roati-PRL02,Ott-PRL04,Silber-PRL05,Gunter-PRL06,Ospelkaus-PRL06,Zaccanti-PRA06,McNamara-PRL06,Best-PRL09,Fukuhara-PRA09b,Deh-PRA10,Tey-PRA10,Sugawa-NP11,Schuster-PRA12,Tung-PRA13,Ferrier-Barbut-S14,Delehaye-PRL15,Vaidya-PRA15,XCYao-PRL16,YPWu-JPB17,Roy-PRL17,Schafer-PRA18}. New phenomena, such as phase separation~\cite{Lous-PRL18} or Bose-Fermi superfluid mixtures~\cite{Trautmann-PRL18}, have been observed in clean and fully controllable setups, where the inter- and intraspecies interactions can be tuned.\par 
To fully comprehend the properties of such mixtures, those of the independent systems need to be well understood. This is indeed the case for several bosonic and fermionic gases. Namely, phase transitions between Mott insulator and gapless states in locally interacting systems have been widely studied for both statistics. It is well known that bosonic systems exhibit Mott insulator phases at integer densities~\cite{Cazalilla-RMP11}, while for two-color fermions this phase emerges only at half-filling~\cite{Guan-RMP13}. When fermions and bosons are mixed, a rich scenario is expected, and different levels of theoretical approach have been considered over the years.\par 
The first approach to describing a mixture of bosons and fermions consists of freezing their internal degrees of freedom, a scenario that has been widely studied~\cite{Albus-PRA03,Cazalilla-PRL03,Lewenstein-PRL04,Mathey-PRL04,Roth-PRA04,Frahm-PRA05,Batchelor-PRA05,Takeuchi-PRA05,Pollet-PRL06,Mathey-PRA07,Mering-PRA08,Suzuki-PRA08,Luhmann-PRL08,Rizzi-PRA08,Orth-PRA09,XYin-PRA09,Sinha-PRB09,Orignac-PRA10,Polak-PRA10,Mering-PRA10,Masaki-JPSJ13}. Among the diverse states  revealed by these studies, we emphasize the insulator phases at integer bosonic densities and the mixed Mott insulator determined by the relation $\rho_B+\rho_F =1$, where $\rho_B$ and $\rho_F$ are the bosonic and fermionic densities respectively~\cite{Zujev-PRA08}. An insulator that fulfills this commensurability relation has been observed in experiments~\cite{Sugawa-NP11}.\par 
To enrich the description, it is necessary to consider internal degrees of freedom, which are relevant for both bosons and fermions. Inspired by the BCS theory, several authors have studied mixtures of two-color fermions and scalar bosons at particular densities, using bosonization~\cite{Mathey-PRL04}, renormalization group~\cite{Mathey-PRL06,Mathey-PRB07,Klironomos-PRL07}, mean-field theory~\cite{Sengupta-PRA07,Anders-PRL12,Bukov-PRB14,TOzawa-PRA14}, and dynamical cluster~\cite{Bilitewski-PRB15} approaches in one, two, and three dimensions. In those studies, diverse ground states were reported, such as superfluid, spin-density wave, charge-density wave (CDW), phase separation, Mott insulator, supersolid, antiferromagnetic order, and evidence of various types of pairing, among other phenomena. In a recent paper, we numerically explored the above model in one dimension  considering the hard-core limit and only repulsive interactions. There we obtained two non-trivial insulators phases that fulfill the relations $\rho_B+\rho_F=1$ and $\rho_B+\tfrac{1}{2}\rho_F=1$ for a fixed fermionic density~\cite{Avella-PRA19}. This indicates that considering the internal degrees of freedom of fermions leads to a new non-trivial insulator, but restricting the Hilbert space of the bosons prevents the emergence of bosonic Mott insulators.\par
Clearly, mixtures of scalar bosons and two-color fermions hide much more phenomena to be discovered. This motivates the present investigation, in which we determine the phase diagrams that emerge when allowing more than one boson per site, i.e. when considering the soft-core approach. This has only been analyzed in a very recent report, where the authors study the FFLO physics in a spin-imbalanced mixture~\cite{Singh-PRR20}.
Taking into account that in cold-atom setups the amplitude and sign of interspecies interactions can be tuned, we considered both repulsive and attractive couplings. Exploring the superfluid-insulator transitions in soft-core mixtures, we  found that regardless of the sign of the boson-fermion interaction and for a fixed fermionic density $\rho_F$, there are always two non-trivial insulator phases between the trivial insulators at integer bosonic densities $\rho_B$.  These satisfy the conditions $\rho_B\pm\rho_F=n$ and $\rho_B\pm \tfrac{1}{2}\rho_F=n$ ($n$ integer), and the plus (minus) sign for repulsion (attraction). Since in experiments the number of fermions can be changed while fixing the density of bosons, we also perform a similar exploration and observe only three non-trivial insulator phases located at densities that fulfill the above conditions, where the missing one leads to a non-physical situation. Our investigation thus establishes the emergence of insulator phases for boson-fermion attraction, which had not been reported until now.\par 
\begin{figure}[t] 
\includegraphics[width=18pc]{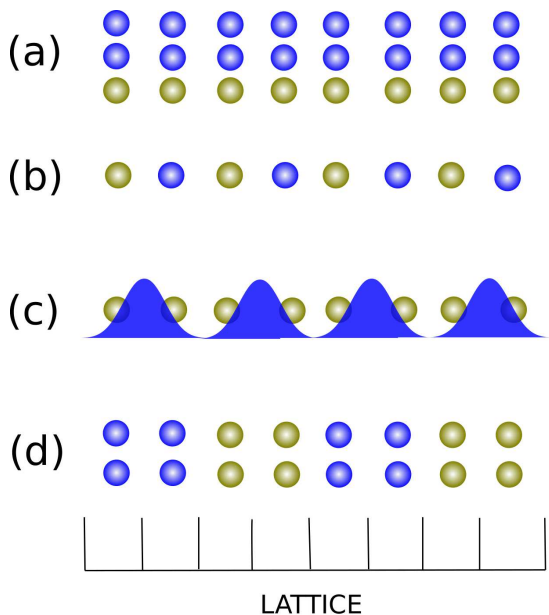} 
\caption{\label{fig1} Illustration of schematic ground states of a mixture of scalar bosons and two-color fermions in one-dimension. Here we consider a lattice with eight ($L=8$) sites and draw different possible distributions of particles.
Blue (golden) circles represents bosons (fermions). (a) Coexistence of Mott insulators for fermions ($\rho_F = 1$) and bosons ($\rho_B = 2$); here $U_{BF} \lessgtr0$. (b) Mixed Mott insulator state with $\rho_F = 1/2$ and $\rho_B = 1/2$ for repulsive interparticle coupling. (c) Noncommensurate insulator state with a fermionic density $\rho_F = 1$ and bosonic density $\rho_B = 1/2$, for $U_{BF} \lessgtr0$. (d) Phase separation state for repulsive interactions, $\rho_F = 1$ and $\rho_B = 1$.}
\end{figure} 
The outline of this paper is as follows. The model used to describe a mixture of bosonic and fermionic atoms is introduced 
in Sec.~\ref{sec2}. The superfluid-insulator transitions and some appropriate relations to locate them are discussed in Secs.~\ref{sec3} and ~\ref{sec4} for repulsive and attractive boson-fermion interactions, respectively. The special case of fermionic half-filling density is discussed in Sec.~\ref{sec5}. A summary of our conclusions is presented in Sec.~\ref{sec6}.\par
\section{\label{sec2} Bose-Fermi-Hubbard Model}
We start by describing the model and the main approaches considered in the current investigation for  studying a degenerate mixture of bosons and fermions.\par 
A system of scalar bosons in one dimension can be modeled by the Hamiltonian 
\begin{equation}\label{BHamil}
    \hat{H}_{B}=-t_B\sum_{\langle i,j\rangle}\left(\hat{b}_{i}^{\dag}\hat{b}_{j}+ \text{h.c.}\right)+\frac{U_{BB}}{2}\sum_{i}\hat{n}_{i}^{B}\left(\hat{n}_{i}^{B}-1\right),
\end{equation}
\noindent which takes into account the kinetic energy (first term) and the local repulsive interaction between bosons (second term). In Hamiltonian \eqref{BHamil}, $\hat{b}_{i}^{\dag}$  ($\hat{b}_{i}$) creates (annihilates) a scalar boson at size $i$. The local boson number operator is $\hat{n}^{B}_{i}=\hat{b}_i^{\dag}\hat{b}_i$. The parameter $U_{BB}$ quantifies the local interaction, and $t_B$ is the hopping amplitude between neighboring sites $(\langle i,j\rangle)$.\par 
A system composed of two-color fermions that interact locally is described by the Hamiltonian
\begin{equation}\label{FHamil}
    \hat{H}_{F}=-t_F\sum_{\langle i,j\rangle\sigma}\left(\hat{f}_{i,\sigma}^{\dag}\hat{f}_{j,\sigma} + \text{h.c.}\right)+\frac{U_{FF}}{2}\sum_{i,\sigma\neq\sigma'}\hat{n}^{F}_{i,\sigma}\hat{n}^{F}_{i,\sigma'},
\end{equation}
\noindent $\hat{f}_{i,\sigma}^{\dag}$ ($\hat{f}_{i,\sigma}$) being an operator that creates (annihilates) a fermion with internal degree of freedom  
$\sigma=\uparrow,\downarrow$ at site $i$. The local operator $\hat{n}^{F}_{i,\sigma}=\hat{f}_{i,\sigma}^{\dag}\hat{f}_{i,\sigma}$ corresponds to the density operator for $\sigma$-fermions. The nearest-neighbor fermionic hopping parameter is $t_F$, and $U_{FF}$ quantifies the fermion-fermion interaction. The fermionic density for systems with two-color fermions varies in the interval $[0,2]$, so that $\rho_F=1$ corresponds to half-filling.\par
When two-color fermions and scalar bosons are mixed in a one-dimensional optical lattice and interact with each other, they are described by the Hamiltonian 
\begin{equation}\label{BFHamil}
    \hat{H}_{BF}=\hat{H}_B+\hat{H}_F+U_{BF}\sum_{i,\sigma}\hat{n}^{B}_{i}\hat{n}^{F}_{i,\sigma},
\end{equation}
\noindent where the boson-fermion interaction $U_{BF}$ can be repulsive or attractive ($U_{BF} \lessgtr0$). We measure energies and gaps in units of the fermionic hopping parameter $t_F$ i.e., we set $t_F= 1$ as the energy scale. From now on, unless stated otherwise, we consider bosonic and fermionic isotopes of the same kind of atoms, hence $t_F=t_B$.\par
Importantly, the number of bosons per site is unbounded, making the local Hilbert space exactly untractable. To deal with the model numerically, it is necessary to perform a cutoff, i.e., we consider the soft-core approximation and restrict the number of bosons per site to a maximum of $\hat{n}_{max}=3$. This results in a large yet tractable local  Hilbert space of dimension $d=16$. Note that it has been argued in several reports that the qualitative physical properties obtained for $\hat{n}_{max}=3$  are unaffected when $\hat{n}_{max}$ is increased~\cite{Pai-PRL96,Rossini-NJP12}.\par 
The ground-state energy $E(N_{\uparrow},N_{\downarrow},N_B)$ for $N_B$ bosons and $N_{\uparrow}$, $N_{\downarrow}$ fermions of a Bose-Fermi mixture described by Hamiltonian~\eqref{BFHamil} is obtained using the density matrix renormalization group  (DMRG) algorithm with open boundary conditions~\cite{White-PRL92,Hallberg-AP06}. We perform several finite-system sweeps until the ground-state energy is converged to an absolute error of $10^{-3}$, keeping a discarded weight of $\sim10^{-7}$ in the dynamic block selection state (DBSS) protocol~\cite{Legeza-PRB03}.\par 
In Fig. ~\ref{fig1}, we sketch some possible distributions of carriers along the lattice, which will emerge depending on the sign of the boson-fermion interaction. For instance, the coexistence of fermionic and bosonic Mott insulator states is depicted in (a), insulator states with commensurate or noncommensurate total number of carriers are shown in (b) and (c), respectively, and an immiscible phase separation state is sketched in (d). Other carrier  distributions can be obtained by varying the densities and interaction parameters, as discussed below.\par 
In addition, we note that the system studied in the present investigation can be implemented in the laboratory. In particular, several mixtures of bosonic and fermionic atoms in a degenerate regime have been achieved in cold-atom setups, even though their stability is severely limited by 3-body recombinations. A promising candidate for emulating  the Hamiltonian \eqref{BFHamil} is a mixture containing $^{174}$Yb and $^{171}$Yb atoms, given that the latter has a nuclear spin $I=1/2$, whereas  the former has zero nuclear spin~\cite{YTakasu-JPSJ09}.\par  
Carrying out a complete study sweeping through all the hopping, interaction and density parameters is a phenomenal task, leading to the several mentioned studies on the Bose-Fermi-Hubbard model. An important conclusion from this theoretical and experimental research is that the sign of the interparticle interaction is highly relevant and determines the response of the mixture. Considering this, we will discuss each type of interaction separately.\par
\onecolumngrid
\begin{figure*}[t]
{\centering
  \includegraphics[width=70mm]{Fig2a.eps}
  \includegraphics[width=70mm]{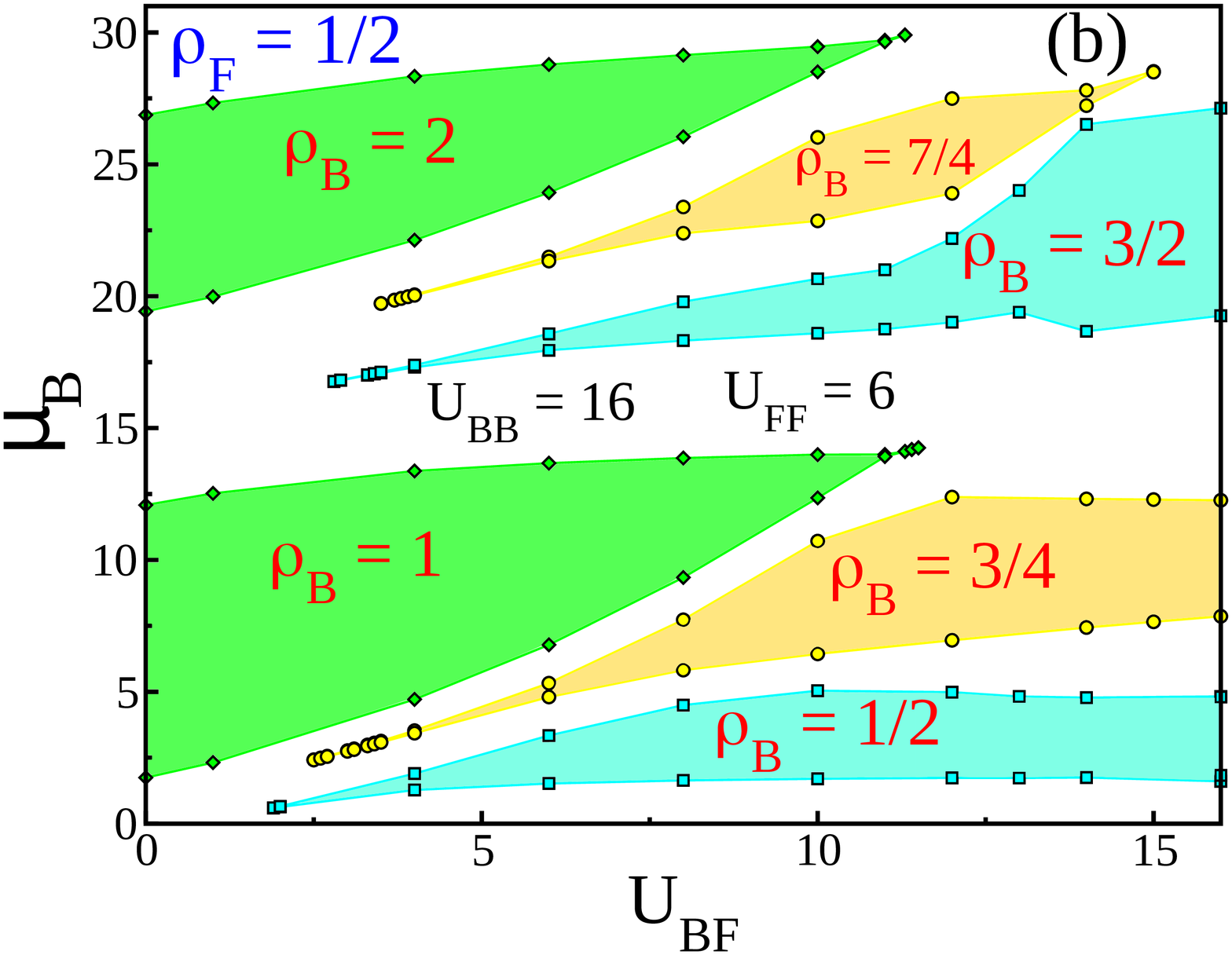}}
\hspace{0.05\textwidth}
\newline
{\centering
  \includegraphics[width=70mm]{Fig2c.eps}
  \includegraphics[width=70mm]{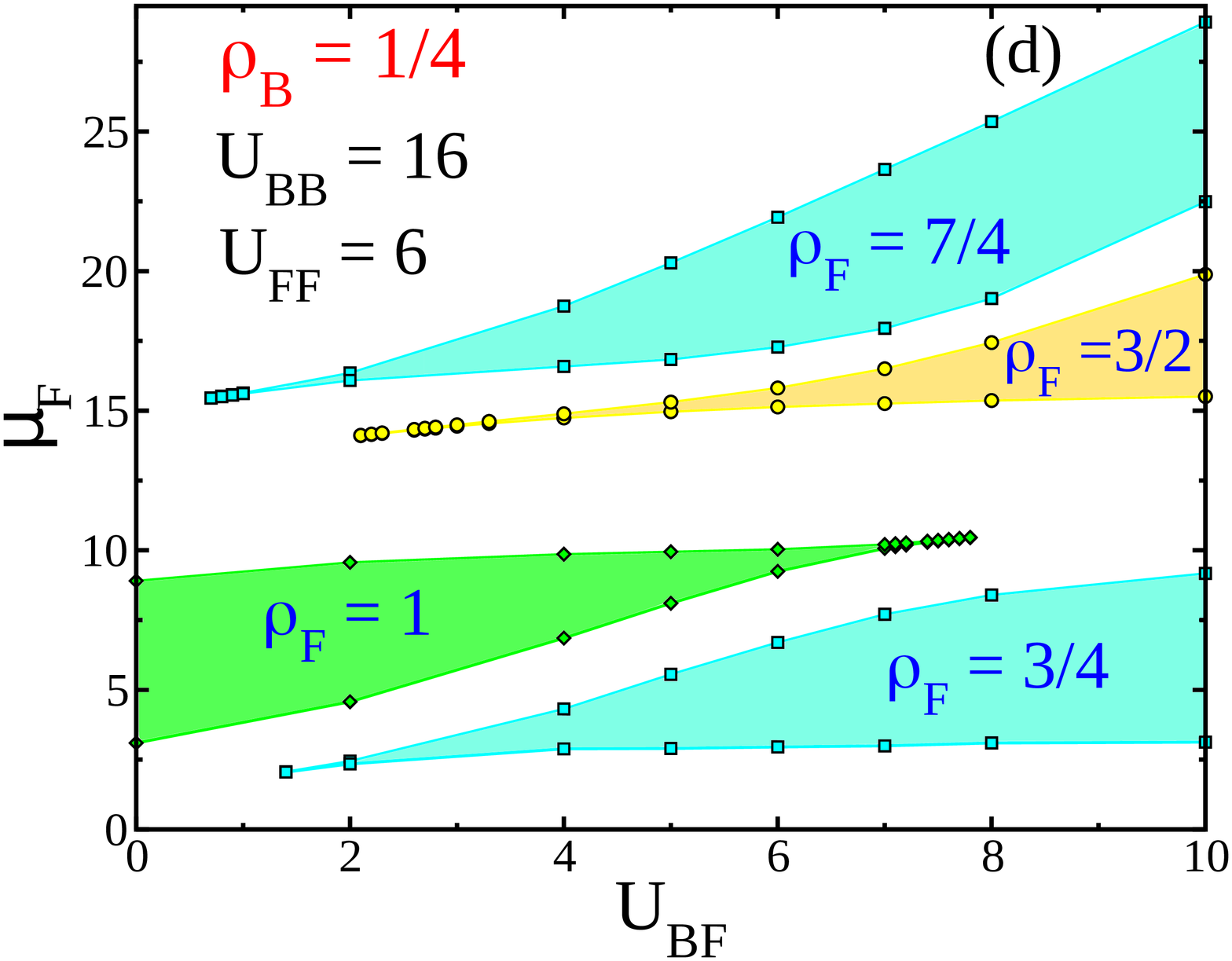}
}
\caption{\label{fig2} Physical properties in the thermodynamic limit of a mixture of scalar bosons and two-color fermions with repulsive boson-fermion coupling. Here the boson-boson and fermion-fermion interactions are $U_{BB}=16$ and $U_{FF}=6$, respectively. (a) Bosonic density ($\rho_B$) versus the bosonic chemical potential ($\mu_B$) for a fixed fermionic density of $\rho_F = 1/2$ and two values of boson-fermion coupling. In the inset, we show the width of the plateaus as a function of the inverse of the lattice size, indicating that they are finite when $1/L\rightarrow 0$ (extrapolated diamond points). (b) Phase diagram in the bosonic chemical potential ($\mu_B$)- interparticle interaction ($U_{BF}$) plane for a fixed fermionic density $\rho_F = 1/2$. The white areas are superfluid regions, while the colored lobes correspond to insulator phases, where the upper (bottom) border is the chemical potential for adding (removing) a boson. (c) Fermionic density ($\rho_F$) versus the fermionic chemical potential ($\mu_F$) for a fixed bosonic density of $\rho_B = 1/4$ and two values of boson-fermion coupling. Again, in the inset we show that the width of the plateaus is finite when $1/L\rightarrow 0$. (d) Phase diagram in the $\mu_F$ vs $U_{BF}$ plane for a fixed bosonic density of $\rho_B = 1/4$. As before, white (colored) areas represent superfluid (insulator) phases. In all the figures, the points correspond to DMRG results and the  lines are visual guides. The values in the thermodynamic limit were obtained by using a second-order polynomial extrapolation.}
\end{figure*}
\hfill{}
\twocolumngrid
%
%
\section{\label{sec3}  Repulsive boson-fermion interaction ($U_{BF}>0$) }
Bose-Fermi mixtures with repulsive interactions have been shown to feature a mixed Mott state that fulfills $\rho_B+\rho_F=1$ for polarized carriers~\cite{Zujev-PRA08}. In addition, when considering an internal structure for fermions, a mixture with hard-core bosons shows the mixed Mott state and a noncommensurate insulator characterized by the relation  $\rho_B+\tfrac{1}{2}\rho_F=1$~\cite{Avella-PRA19}. Hamiltonian~\eqref{BFHamil} goes beyond these cases and describes a mixture of two-color fermions and bosons in the soft-core approximation. Without coupling between fermions and bosons ($U_{BF}=0$) only the well-known Mott insulators (trivial) of each species emerge. Also, we recover the behavior of polarized carriers as the repulsion between fermions is very large ($U_{FF}\rightarrow\infty$), i.e. only the mixed Mott state will appear. In the absence of repulsion between fermions ($U_{FF}=0$), it is difficult to establish the mixed Mott state and the noncommensurate insulators will prevail. Motivated by these findings and intrigued by the possibility of unearthing new properties and characteristics of these mixtures in intermediate scenarios, we consider the more general situation corresponding to the soft-core limit of Eq.~\eqref{BFHamil}.\par
First, fixing the fermionic density at $\rho_F= 1/2$, we increase the number of bosons from zero up to a global density $\rho_B \leq 3$, considering a boson-boson interaction $U_{BB}=16$ and fermion-fermion repulsion $U_{FF}=6$ [see Fig.~\ref{fig2} (a)]. For a weak boson-fermion repulsion of $U_{BF}=1$ (red open squares), the bosonic chemical potential  $\mu_B=E(N_{\uparrow},N_{\downarrow},N_B +1)-E(N_{\uparrow},N_{\downarrow},N_B)$ increases monotonously with the number of bosons, except at integer densities, where large plateaus appear. This is expected from the bosonic limit (without fermions) and the results found for polarized fermions and bosons~\cite{Zujev-PRA08}. Naturally, this is not seen in the hard-core limit~\cite{Avella-PRA19}. For a larger boson-fermion interaction $U_{BF}=8$ (black circles), the trivial plateaus at integer bosonic densities survive, but their width shrinks. Surprisingly, four non-trivial plateaus emerge at the bosonic densities $\rho_B=1/2, 3/4, 3/2,$ and $7/4$. In the inset of Fig.~\ref{fig2} (a), we show the evolution of the width of these plateaus ($\Delta^B=E(N_{\uparrow},N_{\downarrow},N_B +1) +E(N_{\uparrow},N_{\downarrow},N_B -1)-2E(N_{\uparrow},N_{\downarrow},N_B)$)  as the lattice size increases, being finite in the thermodynamic limit. 
Crucially, the plateaus at the bosonic densities $\rho_B=1/2$ and $3/2$ are related to ground states where the total number of the particles (bosons plus fermions) is commensurate with the lattice size, i.e. these insulators correspond to mixed Mott insulators given by the relation $\rho_B+\rho_F=n$, where $n$ is an integer, namely $n=1$ and $2$ for the plateaus at $\rho_B=1/2$ and $3/2$, respectively. On the other hand, the non-trivial plateaus at the bosonic densities $\rho_B=3/4$ and $7/4$ do not fulfill the commensurability condition and instead satisfy the relation $\rho_B+\tfrac{1}{2}\rho_F=n$, recovering the particular bosonic densities with $n=1$ and $2$. The latter non-trivial insulators imply that the number of bosons plus the number of any kind of fermions are  commensurate with the lattice, which was recently evidenced as a limiting case of an imbalanced scenario~\cite{GuerreroS-Arxiv20}. The above discussion, as well as calculations for other fermionic densities (not shown), allow us to conclude that a mixture of two-color fermions and scalar bosons can have two insulator states (one of them commensurate) between trivial (integer density) bosonic insulators.\par 
\begin{figure}[t!]
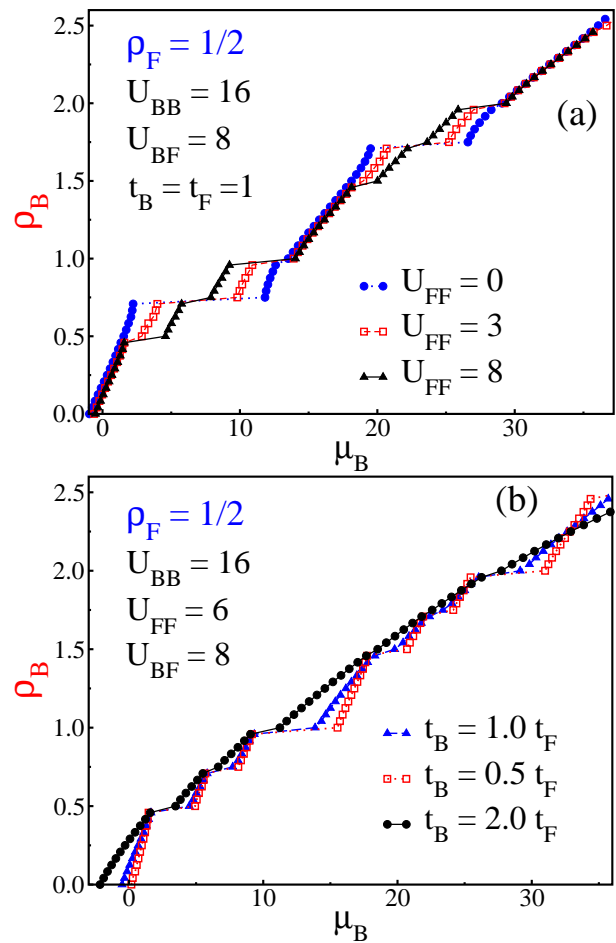

\begin{minipage}{19pc}
\includegraphics[width=19pc]{Fig3a.eps}
\end{minipage}
\hspace{5pc}%
\begin{minipage}{19pc}
\includegraphics[width=19pc]{Fig3b.eps}
\caption{\label{fig3} (a) Evolution of the density versus chemical potential curve for bosons for different values of the fermion-fermion interaction. Here, we consider $t_B=t_F=1$ and $U_{FF}= 0,3$, and $8$. (b) Bosonic density versus bosonic chemical potential for mixtures with mass asymmetry between the carriers. Here, $U_{FF}=6$ and $t_B/t_F=0.5, 1$, and $2$.  
In both plots, the fermionic density is $\rho_F = 1/2$, $U_{BB}=16$, and $U_{BF}=8$. The lines are visual guides.}
\end{minipage}\hspace{2pc}%
\end{figure}
These results clearly indicate that both the fermionic and bosonic densities, as well as their coupling, determine the existence and the properties of the insulating phases. To present a more complete picture, we show a phase diagram of the bosonic chemical potential versus the boson-fermion interaction, keeping constant the fermionic density $\rho_F= 1/2$, the boson-boson interaction $U_{BB}=16$, and the fermion-fermion repulsion $U_{FF}=6$ (see Fig.~\ref{fig2} (b)). 
The colored regions are insulating phases, while the white ones correspond to gapless phases, i.e. superfluid states. The trivial bosonic Mott lobes (green areas) shrink as the repulsive boson-fermion coupling increases, with critical points indicating their suppression at $U^{*}_{BF}\approx 11.7$ and $11.3$ for $\rho_B= 1$ and $2$, respectively. In  our case, the fermion-fermion interaction makes the Mott insulator lobes disappear more quickly than the prediction for a mixture of scalar bosons and polarized fermions, namely $U^{*}_{BF}\approx 2U_{BB}$~\cite{Zujev-PRA08}. In the latter study and in our previous hard-core approach~\cite{Avella-PRA19}, it was shown that the non-trivial lobes emerge from a finite value of the boson-fermion repulsion, a scenario that is seen here in the most general case. Contrary to what is observed in the hard-core limit~\cite{Avella-PRA19}, each mixed Mott lobe (cyan areas) appears earlier than the closest noncommensurate lobe (yellow areas) due to the lower repulsion between bosons. Specifically, the non-trivial lobes for densities $\rho_B=1/2, 3/4, 3/2$, and $7/4$ emerge at the critical points  $U^{*}_{BF}\approx 1.9, 2.5, 2.9$, and $3.9$, respectively. Notice that the width of the mixed Mott lobes tends to saturate for large values of the boson-fermion interaction, showing that this feature does not depend on the boson repulsion. Furthermore, the evolution of noncommensurate lobes differs from the hard-core result, where the width always increases~\cite{Avella-PRA19}. Now we see that for $\rho_B=3/4$, the width saturates for larger values of $U_{BF}$, and vanishes for $\rho_B=7/4$, determining a closed lobe in the phase diagram.\par
\begin{figure}[t] 
\includegraphics[width=18pc]{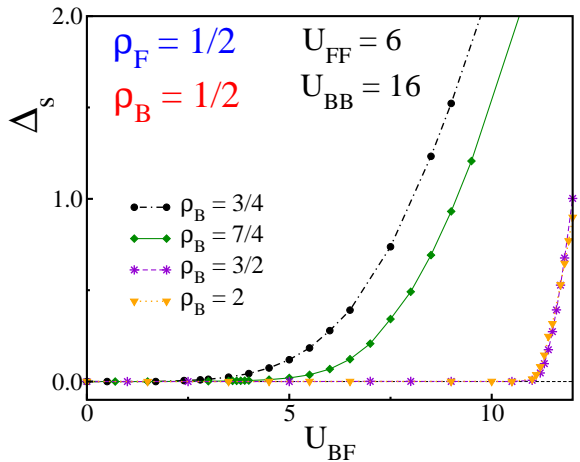} 
\caption{\label{fig4} Spin gap $\Delta_S$ as a function of the boson-fermion interaction $U_{BF}$ for a fixed $\rho_F = 1/2$. For all the considered bosonic densities, a gapless and a gapped region are clearly seen. The lines are visual guides.} 
\end{figure} 
When studying mixtures of bosons and fermions, it is a common practice to fix the fermionic density and vary the number of bosons; however, in experiments both can be controlled. To provide more evidence of the revealed  density conditions, we explore the superfluid-insulator transitions, fixing the bosonic density ($\rho_B=1/4$) and varying the number of fermions. We define the fermionic chemical potential as $\mu_F=E(N_{\uparrow}+1,N_{\downarrow}+1,N_B)-E(N_{\uparrow},N_{\downarrow},N_B)$; its evolution as the number of fermions per site varies from zero to two is shown in Fig.~\ref{fig2} (c). Again we see that for a weak boson-fermion repulsion $U_{BF}=1$ (red squares), the bosons and fermions are quasi-independent (compare to Fig.~\ref{fig2} (a)), and there is only one plateau at half-filling, as expected from the exact solution of the Fermi-Hubbard model (without bosons)~\cite{Lieb68}. Increasing the repulsion between bosons and fermions to $U_{BF}=8$ (black circles), the antiferromagnetic Mott insulator phase disappears, whereas three non-trivial insulating phases emerge at the fermionic densities $\rho_F= 3/4, 3/2$, and $7/4$. We see that the mixed Mott insulator states are present, since the plateaus at $\rho_F= 3/4$ and $7/4$ correspond  to a total number of particles equal to and twice the lattice size, respectively. The remaining plateau $\rho_F= 3/2$ satisfies the relation $\rho_B+\tfrac{1}{2}\rho_F=1$. Here, one non-trivial plateau is missing, because the mathematical relation reported above leads to an unphysical situation (fermionic density $\rho_F=7/2>2$). The charge gap ($\Delta^F=E(N_{\uparrow}+1,N_{\downarrow}+1,N_B) + E(N_{\uparrow}-1,N_{\downarrow}-1,N_B)-2E(N_{\uparrow},N_{\downarrow},N_B)$) for each non-trivial insulating phase as a function of the inverse of the lattice size is shown in the inset of Fig.~\ref{fig2} (c). Using a second-order polynomial extrapolation, we obtained that this gap is always finite; therefore, these insulating phases survive in the thermodynamic limit.\par 
In Fig.~\ref{fig2} (d), we show the corresponding phase diagram in terms of the fermionic chemical potential versus the boson-fermion repulsion for a fixed bosonic density $\rho_B=1/4$ and boson-boson (fermion-fermion) interaction $U_{BB}=16$ ($U_{FF}=6$). As in Fig.~\ref{fig2} (b), the white regions are superfluid, whereas the colored ones correspond to insulator phases. In the absence of boson-fermion interaction, only the trivial Mott insulator phase (green area) emerges, which shrinks as the interaction between fermions and bosons increases, disappearing at $U^{*}_{BF}\approx 7.8$. Also, as the boson-fermion coupling increases from zero, the non-trivial commensurate lobes (cyan areas) emerge at the critical points 
$U^{*}_{BF}\approx 1.4$ and $0.7$ for $\rho_F= 3/4$ and $7/4$, respectively. However, the evolution of these lobes is different; whereas the charge gap for $\rho_F= 3/4$ tends to saturate for larger values of $U_{BF}$, that of $\rho_F= 7/4$ varies, as determined by the increase in chemical potentials for increasing or decreasing the number of fermions. The noncommensurate insulator lobe (yellow area) arises from $U^{*}_{BF}\approx 2.1$ and grows monotonously. We expect that these critical points will take different values as the densities and the other interaction parameters vary.\par 
The influence of the fermion-fermion repulsion on the insulator phases discussed above is depicted in  Fig.~\ref{fig3} (a). Here we consider a Bose-Fermi mixture with a fermionic density of $\rho_F = 1/2$ and parameters $U_{BB}=16$ and $U_{BF}=8$ for boson-boson and boson-fermion interaction, respectively. In the absence of the fermion-fermion repulsion, the noncommensurate insulators dominate, the trivial plateaus at integer densities are narrow, and the mixed Mott insulators do not appear, evidencing the importance of the coupling between fermions for the existence of the latter. Indeed, as  $U_{FF}$ grows, the mixed Mott plateaus emerge and grow as expected, in addition the trivial insulators are also favored. However, the once dominant noncommensurate plateaus decrease with the growth of fermionic repulsion.\par 
Throughout this paper, we consider Bose-Fermi mixture composed of isotopes of the same atom; therefore the assumption $t_B=t_F$ is reasonable. However, we will briefly look at mixtures composed of different atoms and their superfluid-insulator transitions. In  Fig.~\ref{fig3} (b), we show the $\rho_B-\mu_B$ curve for mixtures  with a quarter fermionic filling and hopping parameters $t_B/t_F= 0.5, 1.0$, and $2$. For the set of interaction parameters considered ($U_{BB}=16$,  $U_{BF}=8$, and $U_{FF}=6$) it was found that, as expected, all the insulators decrease when bosons become lighter, a behavior that is more dramatic for bosonic densities greater than 1 where the non-trivial plateaus disappear. These results confirm  that the findings reported here are valid for any kind of mixture and that new features can emerge when the parameters vary even further. From now on, we come back to consider $t_B=t_F=1$.\par
The fermionic particles of our mixture have an internal degree of freedom; therefore, a natural question is whether gapped excitations related to it take place. To explore this issue, we calculate the spin gap $\Delta_S=E(N_{\uparrow}+1,N_{\downarrow}-1,N_B)-E(N_{\uparrow},N_{\downarrow},N_B)$ at each insulator phase. In Fig.~\ref{fig4}, we show the spin gap in the thermodynamic limit as a function of the boson-fermion repulsion for a system with fixed fermionic density $\rho_F=1/2$, $U_{FF}=6$ and $U_{BB}=16$. Note that for this fermionic density, repulsive coupling between fermions and without boson-fermion interaction, we expected a metallic ground state with dominant spin density fluctuations, i.e. both charge and spin gaps vanish~\cite{Penc-PRB94}. Turning on the boson-fermion repulsion and for all the insulating regions, we obtain spin gapless states for a range of values of $U_{BF}$; however, a finite spin gap opens from a critical value, which depends on the bosonic density. We  emphasize that similar results were obtained for attractive boson-fermion coupling and that this unexpected result, which suggests a quantum phase transition in the spin sector, corresponds to an unveiled phenomenon that has not been discussed before in Bose-Fermi mixtures.\par 
\onecolumngrid

\begin{figure*}[t]
\begin{minipage}{13pc}
\includegraphics[width=13pc]{Fig5a.eps}
\end{minipage}\hspace{0.5pc}%
\begin{minipage}{13pc}
\includegraphics[width=13pc]{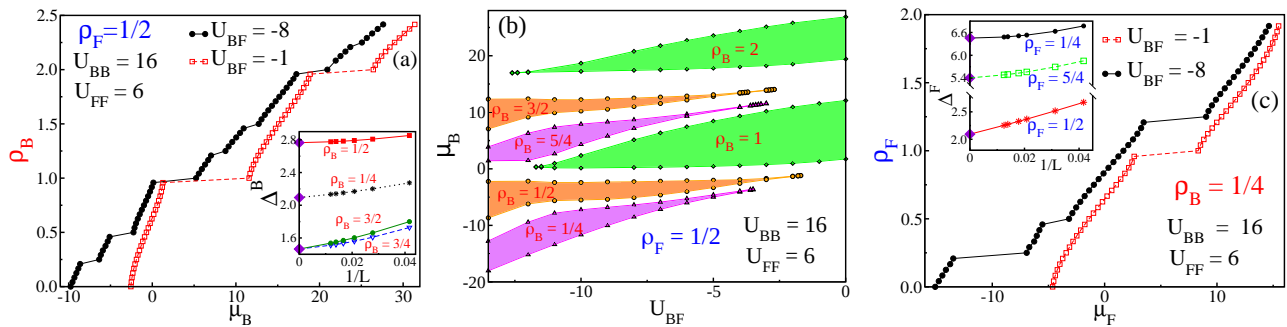}
\end{minipage} \hspace{0.5pc}%
\begin{minipage}{13pc}
\includegraphics[width=13pc]{Fig5c.eps}
\end{minipage} 
\caption{\label{fig5} Quantum phases of a mixture of two-color fermions and scalar bosons for attractive boson-fermion interactions. Repulsive intraspecies interactions were considered ($U_{BB}=16$  and $U_{FF}=6$). (a) Bosonic phase diagram ($\rho_B$ vs $\mu_B$) for a fixed fermionic density $\rho_F = 1/4$ and two values of the boson-fermion coupling. The bosonic charge gap  for each insulator region is shown in the inset for $U_{BF}=-8$ as a function of $1/L$, where extrapolation to the thermodynamic limit is observed. (b) Replicating Fig.~\ref{fig5} (a), we obtain the $\mu_B$ vs $U_{BF}$ phase diagram. The white areas are superfluid regions, while the colored lobes correspond to  insulator phases. (c) Fermionic density profile  $\rho_F$  versus the fermionic chemical potential $\mu_F$ for fixed bosonic density $\rho_B = 1/4$. 
In the inset, we show the evolution of the fermionic charge gap when the lattice grows, showing that it remains finite in the thermodynamic limit. In all the figures, the points correspond to DMRG results and the lines are visual guides.}
\end{figure*}
\hfill{}
\twocolumngrid
\section{\label{sec4}  Attractive boson-fermion interaction ($U_{BF}<0$) }
Attractive interactions between bosons and fermions have been considered by several authors, and interesting effects have been predicted and observed~\cite{Rizzi-PRA08,Mathey-PRL06}. Now, we wish to establish whether the conditions for the emergence of the insulator phases of the mixture change with the nature of the boson-fermion interaction. For this, we maintain the same values of the boson-boson and fermion-fermion couplings considered in Fig.~\ref{fig2} and explore the superfluid-insulator transitions with $U_{BF}<0$; our results are shown in Fig.~\ref{fig5}. First, for a constant global density of fermions $\rho_F = 1/2$, we increase the number of bosons from zero. The corresponding chemical potential is shown in Fig.~\ref{fig5} (a), where attractive boson-fermion interactions of $U_{BF}=-1$ and $U_{BF}=-8$ were considered. Figures ~\ref{fig2}(a) and ~\ref{fig5}(a) have the same parameters,  except that the former is for the repulsive case and the latter for the attractive one; this allows us to clearly see the influence of the nature of the boson-fermion interaction. Again, for weak boson-fermion couplings, only trivial plateaus at integer densities appear, and their widths are independent of the sign of the boson-fermion coupling. The most interesting situation takes place for larger strengths with the emergence of four more plateaus in the bosonic density versus chemical potential curve, as we illustrate for $U_{BF}=-8$ (see Fig.~\ref{fig5}(a)). This confirms that between trivial plateaus, two insulating states arise regardless of the sign of the boson-fermion interaction, this fact being a main conclusion of the present study. For two of the new non-trivial plateaus, namely those at bosonic densities  $\rho_B=1/4$ and $\rho_B=5/4$, the total number of particles is not commensurate with the lattice size. Since the fermionic density is $\rho_F = 1/2$, these new insulator states fulfill the relation $\rho_B-\tfrac{1}{2}\rho_F=n$, where the integer $n$ takes the values $0$ and $1$ for $\rho_B=1/4$ and $\rho_B=5/4$, respectively. These insulator states are characterized by a local coupling between one  fermion and one or more bosons~\cite{GuerreroS-Arxiv20}, forming composite particles, which were evidenced in experiments~\cite{Sugawa-NP11}.  For the other two plateaus, taking place at bosonic densities $\rho_B=1/2$ and $3/2$, the total number of carriers is commensurate with the lattice size, and the condition $\rho_B-\rho_F=n$ ($n$ integer) is satisfied, with $n=0$ and $1$ for $\rho_B=1/2$ and $3/2$, respectively.  However, we note that for other fermionic densities, the plateaus that satisfy the latter condition are such that the total number of carriers is incommensurate with the lattice size. For example, we observed that for $\rho_F=1/3$, the four non-trivial insulating plateaus emerge at the bosonic densities $\rho_B=1/6, 1/3, 7/6$, and $4/3$, none of which satisfy the commensurability relation with the lattice size. Therefore, we are faced with a new scenario, where there is no mixed Mott state.\par  
To illustrate the general behavior of the insulating phases for attractive boson-fermion couplings, a phase diagram in the $\mu_B$ vs. $U_{BF}$ plane is shown in Fig.~\ref{fig5}(b), keeping a fermionic density of $\rho_F = 1/2$ constant  and a boson-boson (fermion-fermion) interaction of $U_{BB}=16$ ($U_{FF}=6$). This phase diagram was obtained by replicating Fig.~\ref{fig5} (a) for several  negative values of $U_{BF}$. The white areas correspond to superfluid regions, which surround the insulator (colored) ones. As in the repulsive case, the trivial lobes shrink and vanish at $U^{*}_{BF}\approx -11.7$ and $-12.6$ for $\rho_B= 1$ and $2$, respectively. A finite value of the boson-fermion coupling is required for non-trivial lobes to arise, determining the critical point located at $U^{*}_{BF}\approx -3.5, -1.7, -3.4$, and $-3.0$ for the bosonic densities $\rho_B=1/4, 1/2, 5/4,$ and $3/2$, respectively.\par 
In Fig.~\ref{fig5}(c), we display the evolution of the fermionic chemical potential as the number of fermions increases, for a mixture with a bosonic density of $\rho_B = 1/4$ and an attractive boson-fermion interaction. This figure corresponds to the attractive version of Fig.~\ref{fig2}(c), and as before, only the antiferromagnetic Mott insulator emerges for small strengths. However, as the boson-fermion coupling increases, the width of this trivial plateau decreases and eventually vanishes, while other non-trivial ones arise at the fermionic densities $\rho_F = 1/4,1/2$, and $5/4$ (the fourth one for $n=-1$ being unphysical, with $\rho_F=5/2>2$). The above fact reinforces our result that the attractive boson-fermion interaction generates  insulator regions different from the repulsive one. Note that the positions of these non-trivial insulator regions fulfill the relations discussed before.\par 
\section{\label{sec5}  Half Filling}
\begin{figure}[t]
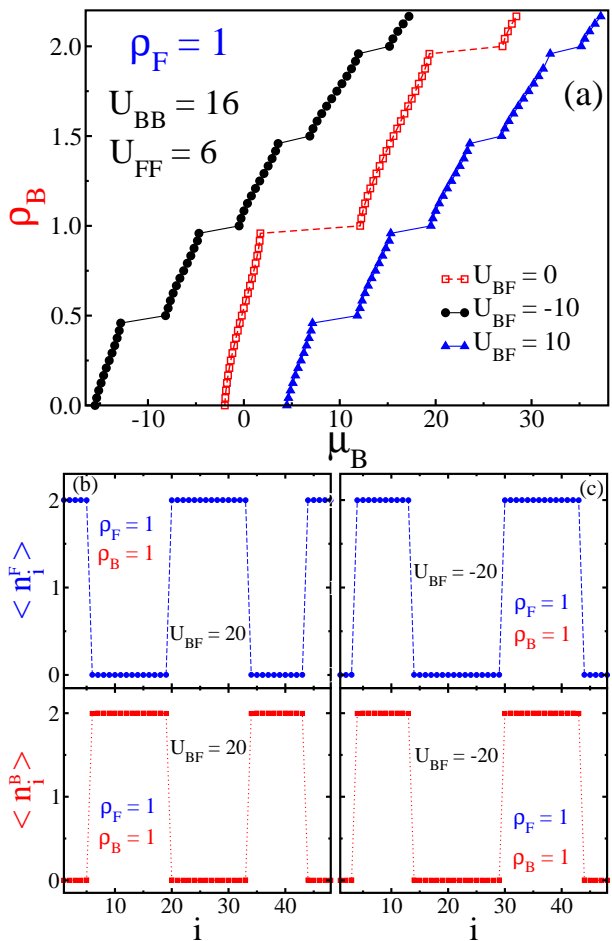

\begin{minipage}{19pc}
\includegraphics[width=19pc]{Fig6a.eps}
\end{minipage}
\hspace{5pc}%
\begin{minipage}{19pc}
\includegraphics[width=19pc]{Fig6b.eps}
\caption{\label{fig6} (a) Bosonic density $\rho_B$ as a function of chemical potential $\mu^B$. The fermion density is $\rho_F = 1$ (half-filling), and the fixed boson-boson (fermion-fermion) repulsion is $U_{BB}=16$ ($U_{FF}=6$). Two different nonzero values of the boson-fermion interaction were considered, namely $U_{BF}=10$ (black) and $U_{BF}=-10$ (red). (b) The distribution of fermions (blue circles) and bosons (red squares) across the lattice for a mixture with repulsive (left panel) and attractive (right panel) interspecies interactions. Here, we consider the same density for bosons and fermions, and equal to one; the other parameters are $U_{BB}=16$, $U_{FF}=6$, and $U_{BF}=\arrowvert20\arrowvert$. In both figures, the lines are visual guides, whereas the points correspond to DMRG results.}
\end{minipage}\hspace{2pc}%
\end{figure}

A case that deserves special attention is that of  half-filling, because it is well known that this configuration leads to interesting physical phenomena in fermionic systems. In the absence of bosons, only at half-filling ($\rho_F = 1$) an  insulator phase is expected, which corresponds to the well-known Mott insulator state, where each site is occupied by one fermion and antiferromagnetic order is established along the lattice. Adding bosons to the system, but without coupling them to the fermions, we trivially expect a superfluid-to-Mott insulator transition under a fermionic Mott background, which takes place when the bosonic density reaches integer values. A coexistence of fermionic and bosonic Mott insulators is thus established for $U_{BF}=0$, as seen in Fig.~\ref{fig6} (a). Turning on the boson-fermion coupling, we observe that the trivial boson plateaus shrink as the interparticle interaction grows, and both will disappear at some large $U_{BF}$. Therefore, the Mott insulator for bosons and fermions coexists in the system for finite values of the interparticle coupling. Regardless of the sign of the boson-fermion interaction, only one non-trivial plateau emerges between the trivial bosonic plateaus, namely at densities $\rho_B = 1/2$ and $3/2$, which agrees with the relations found above for the repulsive and attractive cases.\par 
Finally, we discuss the spatial distribution of particles across the lattice for different states. A homogeneous profile of carriers is obtained in the non-trivial plateaus for weak values of the boson-fermion interaction, which is characterized by one fermion per site and one or three bosons extended across two sites at the densities $\rho_B = 1/2$ and $3/2$, respectively. For larger values of $U_{BF}$, interwoven CDW orderings for bosons and fermions emerge, where the particular form of the density profiles naturally depends on the repulsive or attractive character of the interaction.\par
A different scenario emerges for $\rho_F= \rho_B= 1$, where for weak interspecies interaction the Mott insulator for bosons and fermions coexists, and on average there is one boson and one fermion per site. However, this picture can change, depending on the magnitude and sign of $U_{BF}$. Namely, it can lead to a redistribution of fermions and bosons along the lattice, which opens the possibility for them to occupy the same or different domains of agglutinated particles, i.e., the well-known boson-fermion miscibility problem will arise here~\cite{Mathey-PRL04,Mathey-PRA07,Sugawa-NP11}. For large and positive values of the boson-fermion coupling, fermions and bosons occupy different domains along the lattice, establishing a phase separation state, as sketched in Fig. ~\ref{fig1}(d) and clearly seen in the density profiles shown in Fig.~\ref{fig6} (b) for $U_{BF}=20$. Therefore a quantum phase transition between insulating states of a different nature takes place. A similar effect occurs for attractive interparticle interactions; however, in this case and for large magnitudes of $|U_{BF}|$, bosons and fermions share the same domains (see Fig.~\ref{fig6}(c)), leaving regions of the lattice without particles. In other words, the ground state is a miscible phase separation characterized by domains with or without carriers. It is important to emphasize that recently a phase separation state was observed in a mixture of $^{41}K$ and $^{6}$Li atoms~\cite{Lous-PRL18} using an interspecies Feshbach resonance for controlling the repulsive interaction. It was clearly seen that bosons and fermions occupied different domains in the lattice, as we show in Fig.~\ref{fig6}(b). Thus, it would be possible to identify experimentally the phase separation states predicted by our study, provided that atoms with the correct nuclear spin are used.\par 
\section{\label{sec6} Conclusions}
We studied the ground state of a one-dimensional mixture of scalar bosons and two-color fermions in the soft-core regime, using the density matrix renormalization group technique. Relaxing the hard-core restriction, but keeping up to three bosons per site, we kept the system numerically tractable leading us to unveil new phenomena. For repulsive intraspecies interactions, we swept through a wide range of bosonic and fermionic densities, for positive and negative interspecies couplings, and obtained rich zero-temperature phase diagrams.\par 
Choosing repulsive boson-fermion interactions and fixing the fermionic density $\rho_F$, two non-trivial plateaus arise between the trivial Mott insulators as the number of bosons increase from zero, which satisfy the relations $\rho_B+\rho_F=n$ and $\rho_B+\tfrac{1}{2}\rho_F=n$, $n$ being an integer. As the boson-fermion coupling increases, the non-trivial insulator phases emerge. For stronger couplings, the Mott insulator phases disappear. This generalizes the previous results for polarized atoms and a mixture of two-color fermion and scalar bosons in the hard-core limit~\cite{Zujev-PRA08,Avella-PRA19}. To reinforce our conclusions, we also fixed the bosonic density and varied the fermionic density. Here, we found the trivial antiferromagnetic Mott insulator and three non-trivial insulators that fulfilled the relations given before. In addition, we observed that increasing the fermion-fermion interaction enhances the trivial and mixed Mott insulators, while decreasing the noncommensurate ones. Furthermore, we evidenced that lighter bosons degrade the insulating phases.\par 
For attractive boson-fermion interactions, we observed insulator phases for integer and fractional bosonic densities, where the latter can be commensurate or not with the lattice size; this establishes a fundamental difference from the repulsive case. Therefore, the relations that determine the non-trivial insulator states for attractive interspecies interactions differ from those reported before. Namely, the corresponding relations are given by $\rho_B-\rho_F=n$ and $\rho_B-\tfrac{1}{2}\rho_F=n$, $n$ being an integer. This constitutes one of the main results reported in our paper. We also showed that for a finite interparticle coupling, the trivial Mott insulator states for bosons and fermions coexist in the system. \par 
The energy cost to generate a spin flip in the system was calculated for each insulator phase. We found that the spin gap is zero for a range of values of the boson-fermion coupling and that there is a different critical point for each insulator phase, in which it becomes finite. This suggests a diverse magnetic behavior of the system.\par
Our work motivates the study of Bose-Fermi mixtures in a wide variety of scenarios. For example, in spite of long-range mediated interactions between carriers~\cite{Edri-PRL20,DCZheng-ArX20}, dynamical analysis can be efficiently performed with DMRG methods in systems of particles of different species and statistics~\cite{brockt2015prb,Garttner-PRA19,jj2019prb,stolpp2020prb,jansen-ArX20}. Furthermore, considering that the mixed Mott state (commensurate), phase separation, among other interesting phenomena, have been observed in experiments with bosonic and fermionic isotopes in cold-atoms setups, we expect that our results will stimulate experimentalists to implement the insulator states reported in our investigation.\par 
\section*{Acknowledgments}
R. A. thanks the support of Ministerio de Ciencia, Tecnolog\'{\i}a e Innovaci\'on (MinCiencias)
(Grant No. FP44842-135-2017). J. J. M-A is thankful for the support of Ministerio de Ciencia, Tecnolog\'{\i}a e Innovaci\'on (MinCiencias), through the project \textit{Producci\'on y Caracterizaci\'on de Nuevos Materiales Cu\'anticos de Baja Dimensionalidad: Criticalidad Cu\'antica y Transiciones de Fase Electr\'onicas} (Grant No. 120480863414).
\bibliography{Bibliografia}

\end{document}